\documentstyle[aps,prb,epsfig,floats]{revtex}
\begin{document}
\twocolumn[
\hsize\textwidth\columnwidth\hsize\csname@twocolumnfalse\endcsname
\draft

\title{Magnetic susceptibility of the normal--superconducting transition in high-purity single crystal $\alpha$-uranium}
\author{J. L. O'Brien,$^{1,2}$\cite{email}A. R. Hamilton,$^{2}$
R. G. Clark,$^{2}$ C. H. Mielke,$^{3}$ J. L. Smith,$^{3}$ J. C.
Cooley,$^{3}$ D. G. Rickel,$^{3}$ R. P. Starrett,$^{2}$ D. J.
Reilly,$^{2}$ N. E. Lumpkin,$^{2}$ R. J. Hanrahan, Jr.,$^{3}$
and W. L. Hults$^{3}$ }
\address{$^1$Centre for Quantum Computer Technology, Department of Physics, University of Queensland, Brisbane 4072, Australia}
\address{$^2$Centre for Quantum Computer Technology, School of Physics, University of New
South Wales, Sydney 2052, Australia}
\address{$^3$Los Alamos National Laboratory, Los Alamos, NM 87545, USA}
\date{\today}
\maketitle
\begin{abstract}
We report the first complex ac magnetic susceptibility
measurements of a superconducting transition in very high quality
single crystal $\alpha$-uranium using micro-fabricated coplanar
magnetometers. We identify an onset of superconductivity at
$T$$\approx$0.7 K in both the real and imaginary components of the
susceptibility which is confirmed by resistivity data. A superconducting volume fraction argument, based on a comparison with a calibration
YBa$_{2}$Cu$_{3}$O$_{7 - \delta}$ sample, indicates that
superconductivity in these samples may be filamentary. Our data also
demonstrate the sensitivity of the coplanar
micro-magnetometers, which are ideally suited to measurements in
pulsed magnetic fields exceeding 100 T.
\end{abstract}
]

Since the 1942 discovery of superconductivity in uranium a
coherent picture of the phenomenon in its compounds has been
developed, perturbed only by the identification of heavy fermion
superconductors amongst these materials.\cite{la-aip-43-1} The
nature of superconductivity in elemental uranium, however, has
remained enigmatic, largely due to the difficulty in producing
pure single crystal samples. In its room temperature
$\alpha$-phase, uranium is not a normal bulk superconductor: it
shows a reverse isotope effect, with transition temperatures
increasing with mass squared;\cite{fo-prl-15-860} and competes
with the formation of charge density wave (CDW) states with
transitions at 43, 37 and 23
K.\cite{sm-prl-44-1612,la-prl-81-2978} As the heaviest naturally
occurring element uranium exhibits a CDW state (typically observed
in quasi-one-dimensional materials), is one of very few elemental
type II superconductors, has a crystal structure which is unique
at ambient pressures,\cite{la-aip-43-1} and has a valence shell
configuration which breaks Hund's third rule.\cite{hj-prl-71-1459}

Early magnetic measurements of $\alpha$-uranium showed
superconductivity with critical temperatures ($T_c$'s) ranging
from 0.68-1.3 K for polycrystalline samples.\cite{la-aip-43-1} In
contrast, an upper limit of $T_c$=0.1 K was observed for single
crystals.\cite{fi-jac-213-254} From these data $T_c$ was
understood to decrease with increasing sample
purity.\cite{la-aip-43-1} The absence of a superconducting
signature in corresponding specific heat
measurements\cite{go-pr-152-432,ho-prl-17-694} led to the
suggestion of ``filamentary" as opposed to bulk superconductivity,
where only regions of interconnected filaments exhibit
superconductivity\cite{la-aip-43-1,fi-jac-213-254,go-pr-152-432,ho-prl-17-694,ge-sc-152-755,ga-pr-154-309}
(not to be confused with the use of filamentary in the early
terminology of Type II materials to describe the mixed state).
Pressure studies revealed $\alpha$-uranium to be one of the most
strongly pressure enhanced superconductors, with a $T_c$ rising to
2.3 K at $P$$\approx$1 GPa.\cite{ga-pr-154-309,sm-pr-140-2620}
Specific-heat measurements at these pressures also revealed a
bulk, rather than filamentary, superconducting
state.\cite{ho-prl-17-694} Following these experiments it was
suggested that at $P$=0 strain filaments are produced by the
highly anisotropic thermal expansion of $\alpha$-uranium at low
$T$. Stabilized $\gamma$-U-$X$ alloys ($X$=Pt, Rh, Cr, Mo) also
demonstrated bulk superconductivity, leading to the proposal of an
alternative mechanism in which the filaments consist of impurity
stabilized networks of $\beta$- and $\gamma$-phases of
uranium.\cite{la-aip-43-1,ge-sc-152-755} There were even
references to unpublished transmission electron microscopy images
of the filaments.\cite{ga-pr-154-309} Subsequent calorimetric
studies indicated that $\alpha$-uranium was in fact a bulk
superconductor at $P$=0.\cite{ba-prb-12-4929} It has since been
accepted that superconductivity in $\alpha$-uranium is a bulk
effect, although these results have never been reconciled with the
earlier studies.\cite{la-aip-43-1,fi-jac-213-254} Very recent
measurements on high-purity single crystals are also supportive of
a bulk effect.\cite{la-prb-63-224510} Despite the early intense
efforts a complete picture of the superconducting state in this
unusual material has yet to emerge.

In this communication we present the first complex ac magnetic
susceptibility measurements on single crystal $\alpha$-uranium, of the highest purity yet produced.\cite{mc-jom-49-N7} An
onset to a superconducting state at $T$$\approx$0.7 K is observed,
confirmed by a transition to zero resistivity at $T$$\approx$0.8
K. We also find evidence for
filamentary superconductivity based on a volume fraction
comparison with measurements of a calibration sample of YBa$_{2}$Cu$_{3}$O$_{7 -
\delta}$ (YBCO) that show a clear
signature of the normal-superconducting transition at
$T$$\approx$95 K. The results also suggest that $T_c$ {\it increases} with sample purity, contrary to the earlier body of work, although the details of any filamentary nature may be important. The coplanar micro-magnetometers used in this
work were specifically developed for high sensitivity magnetic
measurements at low $T$. We identify the compatibility of these
devices with the extreme environments of $\mu$s pulsed fields
exceeding 100 T.
\cite{ka-rsi-68-3843,dz-prb-57-14084,ob-prb-61-1584,brooks}

Although zero resistivity is a classic signature of
superconductivity, such measurements cannot distinguish between
bulk and filamentary states because zero resistance is measured
whenever there is a superconducting percolation path. In
contrast, magnetic measurements have historically provided a very
useful probe of superconductivity. In particular magnetic
susceptibility measurements provide information about flux
shielding and can offer insight into the superconducting volume
fraction in non-bulk samples.\cite{go-sst-1-523} The development
of sensitive magnetometers has enabled susceptibility
measurements to be made where effects are slight and on small
samples where signals are weak. Very sensitive superconducting
quantum interference device (SQUID) magnetometers have been
produced,\cite{ke-apl-40-736} but are incapable of operating in
high magnetic fields. Even more sensitive measurements have been
made with cantilever magnetometers,\cite{ha-apl-75-1140} which
are best suited to anisotropic samples, and there is evidence
that they can be used in pulsed magnetic fields with sufficiently
small samples.\cite{na-rsi-68-4061} Lithographically defined
coplanar micro-magnetometers offer high sensitivity, near perfect
compensation of the coils, the possibility of fabricating the
coils directly onto a sample, and the ability to make measurements
in high magnetic fields.

\begin{figure}
\vspace{0.5cm}
\begin{center}
\includegraphics*[width=7.5cm]{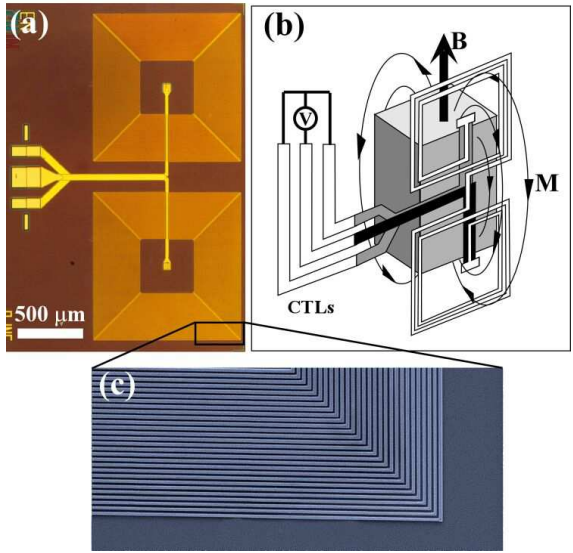}
\end{center}
\caption{Gold coplanar micro-magnetometer and sample mounting
arrangement. (a) Micrograph of 120 turn magnetometer coils (b)
schematic of mounting arrangement of the sample on the
micro-magnetometer coils, showing how the induced magnetization
generates a voltage $V$. (c) Scanning electron microscope image of
the region indicated in (a)} \label{fig1}
\end{figure}

We have designed and micro-fabricated balanced, coplanar coil
magnetometers specifically for magnetic measurements at low $T$
and high $B$. Figure 1(a) shows an optical micrograph of a
magnetometer fabricated on an insulating GaAs substrate using
standard optical lithography techniques. It consists of two
counter-wound coils with a center to center separation of 2mm. The
coils are nearly perfectly compensated because of the precision of
the lithography (Fig. 1c). The magnetometers have been designed to
work with coplanar transmission lines (CTLs) on a printed circuit
board, optimised for ultra-high magnetic field transport
measurements \cite{ka-rsi-68-3843,dz-prb-57-14084,ob-prb-61-1584}
[illustrated schematically in Fig. 1(b)]. The two outer
transmission lines are common and connect to the inner contact of
the upper coil, while the center transmission line contacts the
inner end of the lower coil. This multi-layer design uses
insulating SiN layers to isolate the gold metal interconnects to
the coils, and as a capping layer.

A liquid nitrogen cooled solenoid was used to apply a harmonic
magnetic field $B_{ac}$ parallel to the plane of the magnetometer
coils with a frequency $\nu$=100-150 Hz. This parallel geometry
means that there is no direct coupling between $B_{ac}$ and the
coils, as indicated in Fig. 2b. Since the coils are
counter-wound, any misalignment with respect to $B_{ac}$ will
generate an equal and opposite voltage in each coil. However, if
this parallel magnetic field magnetizes the sample, as indicated
in Fig. 1b, then some of this secondary magnetic flux threads the
two counter-wound coils in opposite senses, producing a voltage
across the coils proportional to $\partial M/\partial t$. The
complex susceptibility $\chi$ was measured by phase sensitive
detection of this voltage. The micro-magnetometers were designed
to maximize the detection of this secondary flux, and those used
in this work had either 80 or 120 turns per coil with a line width
of $\sim$2 $\mu$m.

\begin{figure}
\begin{center}
\includegraphics*[width=8cm]{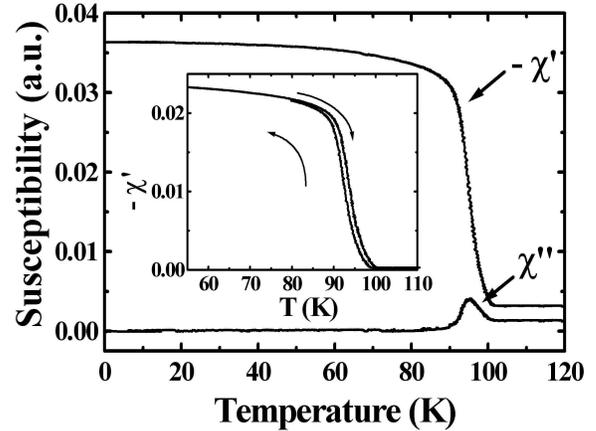}
\end{center}
\caption{Susceptibility vs. temperature for the YBCO sample. Both
-$\chi$$'$ and $\chi$$'$$'$ are shown for $|B_{ac}|$=35 mT and
$\nu$=150 Hz. The inset shows a magnified view of the transition
for up and down temperature sweeps revealing some hysteresis}
\label{fig2}
\end{figure}

\begin{figure}[t!]
\begin{center}
\vspace{0.3cm}
\includegraphics*[width=8cm]{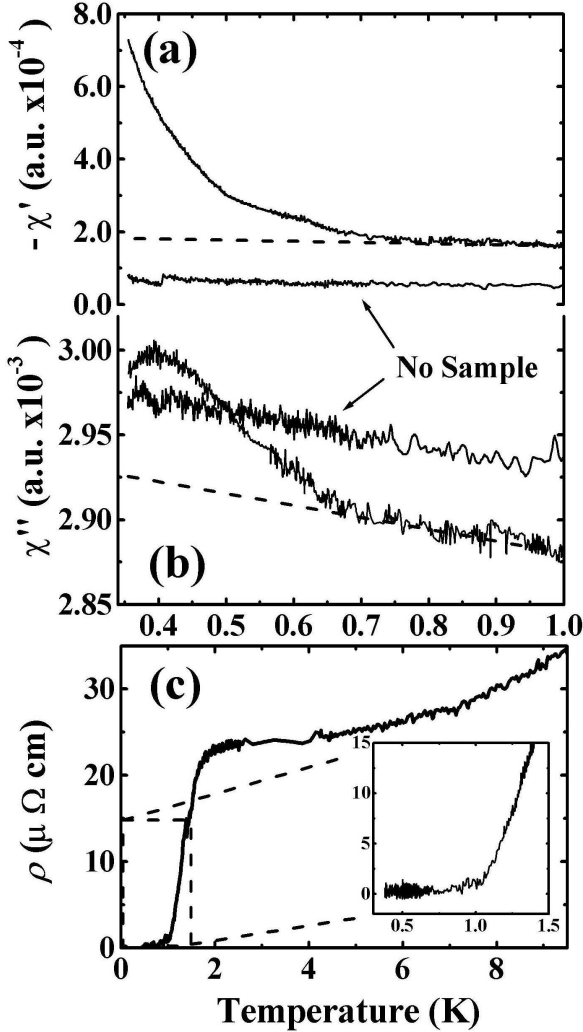}
\end{center}
\caption{Susceptibility vs. temperature for the single crystal
$\alpha$-uranium sample. Both -$\chi$$'$ (a) and $\chi$$'$$'$ (b)
are shown with and without the $\alpha$-uranium sample present.
Data were taken with $|B_{ac}|$=35 mT and $\nu$=150 Hz. Dashed
lines indicate the slope of the ``no sample'' traces. (c) A plot of
resistivity vs. temperature for the same sample. Data have been
interpolated with minimal smoothing for clarity. The inset shows
detail in the transition region.} \label{fig3}
\end{figure}

Two different superconducting samples were used in this study: a
calibration sample of the ceramic high $T_c$ cuprate YBCO; and a
very high quality single crystal $\alpha$-uranium sample. Grains
of YBCO were set in epoxy with the $c$-axes aligned by a magnetic
field and machined into a half cylinder ($r$=0.5 mm). Planar
single crystals of $\alpha$-uranium were grown by
electro-deposition in a salt bath at $\sim$600 $^{\circ}$C with
the $c$-axis perpendicular to the plane.\cite{mc-jom-49-N7} The
residual resisistivity ratio (RRR) $\rho$(300 K)/$\rho$(2 K)
provides a measure of the sample's purity. Resistivity
measurements on this sample show a RRR of 206, three times larger
than any previously reported, indicative of its high purity.
Samples were mounted directly onto the magnetometers and the
magnetometers attached to the CTLs with epoxy. This assembly was
inserted into a $^3$He cryostat giving access to $T$$\geq$0.3 K.


For a superconductor the real component of the susceptibility
$\chi$$'$ is a measure of the magnetic shielding and the imaginary
component $\chi$$'$$'$ a measure of the magnetic
irreversibility.\cite{go-sst-1-523} The signal which is in-phase
with $B_{ac}$ thus measures $\chi$$'$ and the quadrature signal
$\chi$$'$$'$. In order to verify the functionality of the
micro-magnetometers, we first measured the YBCO calibration sample
with $B||c$ (Fig. 2). In the normal state, $T$ $>$ $T_c$, YBCO is
non-magnetic and there is no flux exclusion. Thus for $T$ $>$ 100
K $\chi$$'$ and $\chi$$'$$'$ are both close to zero. As the
temperature is decreased below $T_c$ ($\sim$95 K) supercurrents
are set up to shield the interior of the sample from $B_{ac}$.
This diamagnetic behavior leads to a negative $\chi$$'$ which
becomes more negative as $T$ is reduced and more flux is expelled
from the sample. In this mixed state the flux penetrating the
sample lags the external flux resulting in the dissipation seen in
the $\chi$$'$$'$ signal in Fig. 2. The peak in $\chi$$'$$'$
($T$$\approx$95 K) occurs when the flux is just penetrating as far
as the center of the sample.\cite{go-sst-1-523} At lower $T$ there
is a flux free region at the center of the sample, which becomes
larger as $T$ is decreased further. The dissipation is now
occurring in a smaller fraction of the sample volume and so
$\chi$$'$$'$ now decreases. The inflection point in $\chi$$'$ and
maximum in $\chi$$'$$'$ are the characteristic signatures of a
normal-superconducting transition.\cite{go-sst-1-523} Plots of
$\chi$$'$ near the inflection point for increasing and decreasing
temperature sweeps show a small hysteresis (Fig. 2 inset), in
agreement with established results. These results demonstrate the effectiveness of the micromagnetometers in reproducing known results using an established technique.

\begin{figure}
\begin{center}
\includegraphics*[width=8.5cm]{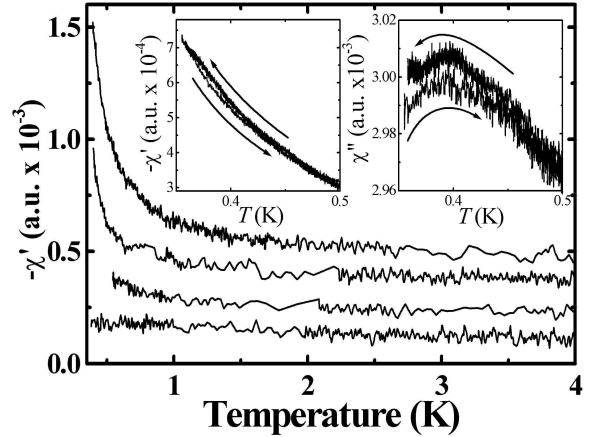}
\end{center}
\caption{Effect of dc magnetic field on magnetic susceptibility
transition in single crystal $\alpha$-uranium. Plots of -$\chi$$'$
with an applied static magnetic field $B_{dc}$ = 0, 2.5, 3.8 and
12.5 mT from top to bottom respectively. All data were obtained
with $|B_{ac}|$ = 35 mT. Traces have been offset for clarity. The
insets shows -$\chi$$'$ and $\chi$$'$$'$ for increasing and
decreasing $T$, indicated by arrows.} \label{fig4}
\end{figure}

Figure 3 shows -$\chi$$'$ and $\chi$$'$$'$ for a single crystal
$\alpha$-uranium sample with an onset to superconductivity at
$T$$\approx$0.7 K. The size of the features are much smaller here
than for YBCO, however, corresponding ``no sample" traces reveal
that the structure is real and not due to the measurement
apparatus. In analogy with the YBCO data, we see a sharp rise in
-$\chi$$'$$'$ ($T$$\lesssim$0.7 K) and a peak in $\chi$$'$
($T$$\approx$0.4 K). In the $\alpha$-uranium case the entire
transition cannot be seen since it is not complete at the lowest
temperature of the $^3$He system. However, the peak in
$\chi$$'$$'$ at $T$$\approx$0.4 K is at the center of the
transition, as for YBCO, and so the data in Fig. 3 represent more
than half of the transition. Figure 3c shows the resistivity
$\rho$ as a function of $T$ for the same sample. The data clearly
show a superconducting transition with an onset at $T$$\approx$1.8
K and zero resistivity point at $T$$\approx$0.8 K. This confirms
that the features in the susceptibility data are due to a
normal-superconducting transition. The value of $T_c$ for this
sample ($\sim$0.8 K) is by far the highest reported for single
crystal $\alpha$-uranium. This is in contrast to the accepted
behavior which suggests that $T_c$ \textit{decreases} with
increasing purity.\cite{la-aip-43-1,note1}

If the transition in $\chi$ is due to superconductivity,
application of a dc magnetic field $B_{dc}$ should move it to
lower $T$. We confirm this by comparing plots of -$\chi'$($T$) for
$B_{dc}$=0, 2.5, 3.8 and 12.5 mT in Fig. 4, which show that the
superconducting behavior is rapidly quenched by a magnetic field.
We note that although only a moderate field is required to
suppress the superconductivity, our observation of a peak in
$\chi$$'$$'$($T$) in Fig. 3 indicates that for $T$$<$0.4 K there
is a flux free region in the sample. This confirms that the
smaller features in -$\chi$$'$ and $\chi$$'$$'$ for
$\alpha$-uranium (Fig. 3) compared to YBCO (Fig. 2) are not due to
penetration of a too large $B_{ac}$ through the whole sample.
Measurements for increasing and decreasing $T$ near the transition
reveal that hysteresis effects in -$\chi$$'$ and $\chi$$'$$'$ are
close the noise limit (Fig. 4 insets). We have also examined the
frequency dependence of this transition and find no measurable
effect over the range $\nu$=100-150 Hz (not shown).

While there has been some controversy surrounding claims of bulk
superconductivity based on susceptibility, it is widely accepted
that these measurements can be used to estimate the
superconducting volume fraction.\cite{go-sst-1-523} If the YBCO
and $\alpha$-uranium samples had identical geometries, a direct
comparison of flux exclusion could be made by comparing the size
of the transition features in $\chi$$'$, using the arbitrary units
which are the same in Figs. 2 and 3. Given that the sample
dimensions are comparable and YBCO is a bulk superconductor, we
estimate that the $\alpha$-uranium excludes a flux equivalent to
$\sim$1\% of the sample volume. While this is a fairly crude
estimate, the difference in transition heights for the two samples
is more than two orders of magnitude, and so clearly significant.\cite{note2}
The London penetration depth $\lambda_L$ can affect the inferred
superconducting fraction,\cite{go-sst-1-523} but it cannot account
for the much smaller transition observed here.

The conclusion that superconductivity in $\alpha$-uranium is
filamentary was dispelled by heat capacity measurements which
revealed bulk superconductivity in {\it polycrystalline}
samples.\cite{ba-prb-12-4929} However, the results presented here on {\it single crystal} 
samples suggest that the superconducting state is filamentary, based on
the volume fraction arguments above. The polycrystalline
result\cite{ba-prb-12-4929} may in fact be due to strain at grain
boundaries ($\alpha$-uranium has highly anisotropic coefficients of thermal expansion\cite{la-aip-43-1}), giving rise to a similar bulk effect as induced at high
$P$.\cite{ga-pr-154-309,sm-pr-140-2620} Impurity effects have been proposed as a mechanism for
filaments in $\alpha$-uranium,\cite{la-aip-43-1,ga-pr-154-309} but
these should be negligible in our high-purity sample. Strain
arising from the anisotropic thermal expansion has also been
suggested,\cite{la-aip-43-1,ge-sc-152-755} however this should not
be relevant in these single crystals.\cite{la-prb-63-224510} Indeed, a Debeye temperature
$\theta_D$=256 K, close to the value of 250 K obtained
from elastic constant measurements suggests that the lattice is
strain free, in contrast to polycrystalline samples.\cite{la-prb-63-224510} A more exotic explanation is
that the distortions in the crystal lattice due to the CDW state
are somehow responsible for causing superconducting filaments.
Resistivity measurements on these samples show clear signatures of
the CDW transitions at 43, 37, and 22
K.\cite{la-prb-63-224510,sm-jsinm-13-833,sm-2001}

The coplanar micro-magnetometers described here are compatible
with the extreme environment of $\mu$s pulsed magnetic fields,
required for future low $T$ de Haas-van-Alphen measurements of
$\alpha$-uranium and high-$T_c$ superconductors such as YBCO. We
have previously demonstrated the capability to make electrical
transport measurements in ms pulsed fields $>$ 50 T\cite{br-prb-59-2604} and $\mu$s pulsed fields $>$ 100 T using the
CTL and sample mounting technology used
here.\cite{ka-rsi-68-3843,dz-prb-57-14084,ob-prb-61-1584,brooks} The CTLs
were specifically designed to eliminated d$B$/d$t$ pickup and the
absence of connecting wires to the magnetometers makes this system
ideally suited to such an environment. Previous de Haas-van Alphen
measurements on LaB$_6$ and CeB$_6$ in ms pulsed magnetic fields
$>$50 T (Ref. \onlinecite{cl-mega-1999}) support this, while the
present work demonstrates extremely sensitive measurements using
these coplanar micro-magnetometers.

In summary, these results represent the first measurements of the
complex magnetic susceptibility of a superconducting transition in
high-purity single crystal $\alpha$-uranium. They suggest that
$T_c$ increases with purity, and indicate that the superconducting
state may be filamentary. This has not been reconciled with recent
results\cite{la-prb-63-224510} and further calorimetric
measurements to lower $T$ are required to resolve this issue. Two
outstanding questions in the $\alpha$-uranium picture of
particular interest are how superconductivity and the CDW states
coexist, and a complete understanding of the CDW state itself.
The high purity single crystal samples and coplanar
micro-magnetometers reported here offer a promising route to
answering these questions. This will require a mapping of the
Fermi surface to determine why particular values of the
wavevector are favourable for the formation of a CDW
state.\cite{la-aip-43-1}
\\

We would like to thank J. C. Lashley and K. -H. M\"{u}ller for
useful discussions. This work was supported by the Australian
Research Council. Work at Los Alamos was performed under the
auspices of the US Dept. of Energy. The National High Magnetic
Field Laboratory is supported by the National Science Foundation.



\begin{thebibliography}{10}

\bibitem[*]{email}
electonic mail: job@physics.uq.edu.au.

\bibitem{la-aip-43-1}
G.~H. Lander, E.~S. Fisher, and S.~D. Bader, Adv. Phys. {\bf
43},    (1994) and Refs. therin.

\bibitem{fo-prl-15-860}
R.~D. Fowler {\it et~al.}, Phys. Rev. Lett. {\bf 15},  860
(1967).

\bibitem{sm-prl-44-1612}
H.~G. Smith, N. Wakabayashi, W.~P. Crummett, and R.~M. Nicklow,
Phys. Rev.
  Lett. {\bf 44},  1612  (1980).

\bibitem{la-prl-81-2978}
L. Fast {\it et~al.}, Phys. Rev. Lett. {\bf 81},  2978  (1998).

\bibitem{hj-prl-71-1459}
A. Hjelm, O. Eriksson, and B. Johansson, Phys. Rev. Lett. {\bf
71},  1459
  (1993).

\bibitem{fi-jac-213-254}
E.~S. Fisher, J. Alloys Compounds {\bf 213/214},  254  (1994) and Refs. therin.

\bibitem{go-pr-152-432}
J.~E. Gordon {\it et~al.}, Phys. Rev. {\bf 152},  432  (1966).

\bibitem{ho-prl-17-694}
J.~C. Ho, N.~E. Phillips, and T.~F. Smith, Phys. Rev. Lett. {\bf
17},  694
  (1966).

\bibitem{ge-sc-152-755}
T.~H. Geballe {\it et~al.}, Science {\bf 152},  755  (1966).

\bibitem{ga-pr-154-309}
W.~E. Gardner and T.~F. Smith, Phys. Rev. {\bf 154},  309  (1967).

\bibitem{sm-pr-140-2620}
T.~F. Smith and W.~E. Gardner, Phys. Rev. {\bf 140},  1620
(1965).

\bibitem{ba-prb-12-4929}
S.~D. Bader, N.~E. Phillips, and E.~S. Fisher, Phys. Rev. B {\bf
12},  4929
  (1975).

\bibitem{la-prb-63-224510}
J.~C. Lashley {\it et~al.}, Phys. Rev. B {\bf 63},  224510
(2001).

\bibitem{mc-jom-49-N7}
C.~C. McPheeters, E.~C. Gay, P.~J. Karell, and J.~P. Ackerman, JOM
{\bf 49},
  n7 22  (1997).

\bibitem{ka-rsi-68-3843}
B.~E. Kane {\it et~al.}, Rev. Sci. Instrum. {\bf 68},  3843
(1997).

\bibitem{dz-prb-57-14084}
A.~S. Dzurak {\it et~al.}, Phys. Rev. B {\bf 57},  14084  (1998).

\bibitem{ob-prb-61-1584}
J.~L. O'Brien {\it et~al.}, Phys. Rev. B {\bf 61},  1584  (2000).

\bibitem{brooks}
J.~S. Brooks {\it et al.}, in {\it Proceedings of the Eighth International Conference on Megagauss Magnetic Field Generation and Related Topics, Tallahassee, 1998}, edited by H. Schneider-Muntau (World Scientific, Singapore).

\bibitem{go-sst-1-523}
F. G{\"o}m{\"o}ry, Supercond. Sci. Technol. {\bf 10},    (1997),
and references therin.

\bibitem{ke-apl-40-736}
M.~B. Ketchen and J.~M. Jaycox, Appl. Phys. Lett. {\bf 40},  736
(1982).

\bibitem{ha-apl-75-1140}
J.~G.~E. Harris {\it et~al.}, Appl. Phys. Lett. {\bf 75},  1140
(1999).

\bibitem{na-rsi-68-4061}
M.~J. Naughton {\it et~al.}, Rev. Sci. Instrum. {\bf 68},  4061
(1997).

\bibitem{note1}
A strong dependence on sample purity was a clue to the pairing mechanism in Sr$_2$RuO$_4$ [A.~P. Mckenzie {\it et~al.}, Phys. Rev. Lett. {\bf 80}, 161 (1998)] and in UPt$_3$ [Y. Dalichaouch {\it et~al.}, Phys. Rev. Lett. {\bf 75}, 3938 (1995)].

\bibitem{note2}
The broad nature of this transition might be related to the details of the superconducting filaments.

\bibitem{sm-jsinm-13-833}
J.~L. Smith {\it et~al.}, J. Supercon.: Incorp. Nov. Mag. {\bf
13},  833
  (2000).

\bibitem{sm-2001}
J.~L. Smith {\it et~al.},   (2001). to appear

\bibitem{br-prb-59-2604}
J.~S. Brooks {\it et~al.}, Phys. Rev. B {\bf 59}, 2604 (1999).

\bibitem{cl-mega-1999}
R.~G. Clark {\it et~al.},   (1998), proc., 8th Intl. Conf. on
Megagauss
  Magnetic Field Generation and Related Topics, Oct. 1998, Tallahassee, USA.

\end{thebibliography}
\end{document}